# Emergent behavior in agent networks: Self-Organization in Wasp and Open Source Communities

Sergi Valverde

ICREA-Complex Systems Lab, Pompeu Fabra University, Dr. Aiguader 80, 08003 Barcelona, Spain
e-mail: svalverde@imim.es

Guy Theraulaz

Centre de Recherches sur la Cognition Animale, CNRS UMR 5169, Université Paul Sabatier, 118 route de Narbonne, 31062 Toulouse Cedex 4, France

Jacques Gautrais

Centre de Recherches sur la Cognition Animale, CNRS UMR 5169, Université Paul Sabatier, 118 route de Narbonne, 31062 Toulouse Cedex 4, France

Vincent Fourcassié

Centre de Recherches sur la Cognition Animale, CNRS UMR 5169, Université Paul Sabatier, 118 route de Narbonne, 31062 Toulouse Cedex 4, France

Ricard V. Solé

(1) ICREA-Complex Systems Lab, Pompeu Fabra University, Dr. Aiguader 80, 08003 Barcelona, Spain
(2) Santa Fe Institute, Hyde Park Road 1399, NM, 87501, USA

**Abstract**

Understanding the complex dynamics of communities of software developers requires a view of such organizations as a network of interacting agents involving both goals and constraints. Beyond their special features, these systems display some overall patterns of organization not far from the ones seen in other types of organizations, including both natural and artificial entities. By looking at both software developers and social insects as agents interacting in a complex network, we found common statistical patterns of organization. Here, simple self-organizing processes leading to the formation of hierarchies in wasp colonies and open source communities are studied. Our analysis further validates simple models of formation of wasp hierarchies based on individual learning. In the open source community, a few members are clearly distinguished from the rest of the community with different reinforcement mechanisms.

**Introduction**

In both nature and engineering, complex designs can emerge as a result of distributed collective processes. In such cases, the agents involved (such as social insects or humans) only have limited access to the global pattern being developed. Of course, an important difference emerges when dealing with the knowledge and goals of the agents themselves. Social insects, for example, have no clue of what they are doing: the overall structure and the final function is not part of the individual knowledge. Instead, engineers belonging to a large team will all know the motivation and final goals. And yet, when the complexity of a given project is large enough, real knowledge at the individual level rapidly shrinks, decisions become largely local and are constrained by how other parts of the project develop. To a large extent, the resulting constraints canalize the possible decisions to be made. And ultimately this means that the possible rules of system construction (at least at some scales) become highly limited. As discussed here, these limitations effectively drive the dynamics of systems growth and development in a way not too different from what we observe in social insect colonies. This is the case of open source software (OSS) development communities.

The reason of comparing such apparently divergent systems is that complex networks of interacting agents often display common patterns of organization. Such type of regularities is not uncommon: they actually reflect the existence of common principles of organization [1] not too different from the ones seen in nature [2].

Although social insect colonies involve simple agents and limited amount of communication among them, they are similar to other communities dealing with complex agents. In trying to understand the origins of the global behavior displayed by programmers inside OSS systems, it can be useful to know if some patterns might be a consequence of simple rules shared by both types of systems. If common patterns are found to exist, we would have evidence for basic principles of self-organization that apply to both insects and humans. As shown below, such common patterns are found in both types of systems.

**Social Networks**

This paper is a comparative study of how social organization takes place in insect and human societies by studying network models of social structure. Social network analysis represents agent relationships with nodes and links [4]. Every node $i$ represents an actor $i$ within the network and links ($i, j$) denote social ties between agents $i$ and $j$. More representative models of social networks decorate each link ($i, j$) with the strength of the social tie [5] or the amount of information flowing through it, hereafter called link weight $w_{i,j}$. The statistical analysis of link weights $w_{i,j}$ between pairs of vertices in the social network indicates an heterogeneous pattern of interactions, typically following a power law:

$$P(w_{i,j}) \sim w_{i,j}^{-\gamma}$$

This means that a few ties are exploited with frequency orders of magnitude larger than many other social ties. However, it has been shown that weak ties enable the fast propagation of information in a social network [5]. In addition, the heterogeneous distribution of link weights might be related to the hierarchical organization of the social network.

The previous modelling approach enables us to make quantitative comparison between human and natural societies and to understand what are the general principles of organization behind them and their differences. To proceed to such comparison we have chosen two specific examples for which it is possible to reconstruct the social network and to properly measure link weights. In addition, both systems display similar patterns of global organization, such as hierarchies and a clear division of labour. Actually, both systems are shown to define networks of interacting agents with very similar features in common, reflecting the presence of limitations in the information shared by agents. As far as we know, this is the first research study concerning the patterns and functional significance of weighted interaction networks in both types of system. One of the main objectives of the present work is to understand to what extent self-organization is responsible for such patterns (including hierarchical structure).

**Weighted Network Data**

We have collected network data from a series of experiments on wasp colonies (see box) [6]. In the social network of a wasp colony, nodes identify individual wasps and links represent hierarchical wasp interactions. Link weight $w_{i,j}$ indicates the number of dominances of wasp $i$ over wasp $j$. A restriction of these experiments is the limited colony size. There is, however, enough data to observe significant statistical correlations (see below).

We also consider the social network of open source software communities. Popular metaphors like "the Cathedral and the Bazaar" suggest that the distributed and unplanned nature of open source development outperforms planned schemes, like proprietary software development [7]. It has been argued that decentralization leads to a distinctive organization hat solves the communication bottleneck associated to large software projects. The amount of submitted e-mails from one programmer to other members is a good indicator of his social position in the software community. However, not every e-mail message has the same influence in the process of software development. In order to reduce the amount of noise, here we will consider only e-mail traffic associated to bug-fixes and bug reporting. The rest of e-mails are discarded from any further consideration. From this subset of e-mails we can reconstruct the social network of the software community as shown in [9].

Nodes and links ($i$, $j$) of the OSS social network represent members and e-mail communication from $i$ to $j$, respectively. At any time, a new software bug is discovered by the member $i$ who sends a notification e-mail. Then, other expert members investigate the origin of the bug and eventually reply with the solution. Typically, several messages are required to solve the problem. Here, we define $E_{i,j}(t)=1$ if developer $i$ replies to developer $j$ at time $t$, or $E_{i,j}(t)=0$ otherwise. We also define link weight $w_{i,j}$ as the amount of e-mail traffic flowing from member $i$ to member $j$:

$$w_{i,j} = \sum_{t=0}^{T} E_{i,j}(t)$$

where $T$ is the timespan of software development. The full dataset of social software networks analyzed here [9] describes the e-mail activity of 120 different software teams from Sourceforge, a large repository of open source projects (http://sourceforge.net). Communities in this dataset range from the very small networks (i.e., one or two members) to the very large (on the order of thousands of members).

In figure 1 we compare social networks from wasp colonies and software communities, with an emphasis in the link weight distributions $P(w_{i,j})$. Here $P(w_{i,j})$ is defined as the probability of having a link with weight $w_{i,j}$. For example, fig.1A displays a social network for a single experiment in a colony of 13 wasps. In order to reduce the noise in our statistical data, we make use of the cumulative distribution $P_>(w_{i,j})$, defined as

$$P_>(w_{i,j}) \sim \int_{w_{i,j}}^{\infty} P(\omega)d(\omega)$$

For the standard case (to be found here) where a scaling behaviour $P(w_{i,j}) \sim w_{i,j}^{-\gamma}$ is observed, we have $P_>(w_{i,j}) \sim w_{i,j}^{-\gamma+1}$.

There is a characteristic pattern of asymmetric interaction, where a few strong wasps dominate the activity of the colony [6]. A similar pattern can be observed in fig.1B for the social network of a small software community. Beyond this qualitative comparison we can find significant agreement in the distribution of link weights. To enable a quantitative comparison between human and wasp societies we have considered the aggregated link weight distribution $P(w_{i,j})$ in an ensemble of 12 small open source communities with an average of 10 programmers each. In spite of the obvious differences between the underlying communities and their small size, the comparison of fig.1C and fig.1D reveals a significant convergence in link weight distributions. Interestingly, the distribution of link weights in large software communities also follows a power-law; with an exponent consistent with the observed in the small software communities (see fig.2C).

**Measures of Centrality: Strength and Out-degree**

There are limitations to what can be understood solely from the analysis of the link weight distribution. A more informative picture of the system emerges from measurements of node importance or centrality [3]. For example, the dominance index of a wasp, computed as the ratio of the number of dominances (DOM) over the total number of hierarchical interactions (DOM+SUB) [4], gives a very reliable image of the wasp's hierarchical rank [6]. It has been conjectured that many successful open source projects also display a hierarchical or onion-like organization. In many of these communities there is a core team of members who contribute most of the code and oversee the design and evolution of the project (see fig. 2D). We can identify these core developers by assuming that members with many social ties have leadership roles in the community. Previous centrality studies of software communities [9] have focused in the node out-degree $k_i$, or the number of social ties outgoing from $i$. However, the out-degree may overlook important (while relatively isolated) members connecting separated sub-teams. In this context, a useful centrality measure for weighted networks is node strength $s_i$ [8],

$$s_i = \sum_{j=1}^{N} w_{i,j}$$

which in software communities equals the total number of e-mails sent by developer $i$. In the wasp colony, node strength coincides with the number of dominances (DOM) and thus it is related to the dominance index employed in biological studies of animal hierarchies.

The distribution of programmer strength in software communities follows a power-law $P(s) \sim s^{-\alpha}$ (see fig. 2A). We can further investigate the origin of this power-law by measuring the dependence of node strength $s$ with out-degree $k$ [8],

$$s(k) \sim k^{\beta}$$

When the exponent $\beta = 1$ there is an absence of correlations between strength and out-degree, that is, link weight $w_{i,j}$ is independent of $i$ and $j$. In this case, both the out-degree and strength are equivalent measures and provide exactly the same centrality information. However, node strength is a better centrality measure in software communities because the observed $\beta$ exponent is significantly larger than 1 (see fig. 2B).

**Self-organization in social networks**

Very simple models of self-organization of hierarchies in animal societies rely on a basic positive feedback mechanism, where successful individuals are reinforced by a simple multiplicative rule [6]. Similarly, we can define the probability of e-mail interaction in a software community as a function of the total number of messages sent by the interacting developers. As the number of messages sent increases so does the likelihood of interaction. Interestingly, in many real weighted networks, the weight $w_{i,j}$ of a link scales with the product of the out-degrees $k_i k_j$ of nodes at its ends [8]. In these systems we measure the dependency of average link weight with $k_i k_j$,

$$\langle w_{i,j} \rangle \sim (k_i k_j)^{\theta}$$

It was found that $\theta \approx 1/2$ for the world-wide airport network [8] and in the metabolic network of *Escherichia coli*. The $\theta$ exponent can be related to the $\beta$ exponent (see previous section). Assuming there are no topological correlations between out-degrees of connected vertices, $\beta = 1 + \theta$ [8]. Then, in uncorrelated networks $\beta = 1$ and $\theta = 0$. By measuring the $\theta$ exponent in our datasets we found $\theta > 0$ exponent and thus giving empirical evidence of a reinforcement mechanism in the experiments of the small wasp colony and in a large software community (see fig. 3). This observation is consistent with the $\beta > 1$ exponent measured in the scaling of strength with out-degree. A more careful analysis reveals the different nature of feedbacks in wasp and human hierarchies.

The comparison between fig. 3B and fig. 3D suggests different mechanisms. In wasp colonies, a simple scaling law of the product of individual tendencies explains the average link weight. In order to reduce fluctuations and to better capture the scaling exponent, we have repeated the least squares fitting with logarithmically binned data. For the wasp dataset we have measured $\theta = 0.36$ exponent (fig. 3B), which is consistent with the exponent $\theta = 0.39$ obtained from the raw dataset (fig. 3A). This simple hypothesis does not fit well the software data, which shows strong nonlinearities. The analysis of logarithmically binned data (see fig. 3D) shows there is an almost flat behaviour for roughly two orders of magnitude and a strong deviation with large $k_i k_j$, which is difficult to notice in the raw dataset (fig. 3C). This deviation is clear for at least two orders of magnitude and thus, unlikely to result from noisy fluctuations in the data. This is a characteristic pattern in many software communities. This clear deviation suggests a pronounced reinforcement effect between strongest members of the community, i.e. core developers (see fig. 2D), which have the largest out-degrees and node strengths.

**Conclusions**

The weighted network analysis of wasp colonies and open source communities revealed the existence of common statistical patterns of social organization. The exploration of communities of interacting systems requires the structural analysis of weighted networks. This is an open area of research that requires more attention. The current framework could also play a key role in order to determine the mechanisms behind social self-organization. In this context, we have provided further empirical evidence of simple self-organization processes leading to the formation of wasp hierarchies. However, sharing collective properties does not imply the underlying organizations comprise the same interaction mechanisms. The same kind of analysis reveals intrinsic differences between wasp and human societies. In the open source community, a few core members are distinguished with reinforcement mechanisms. Arguably, these members might be qualitatively different from other community members. It is future work to select the appropriate type of model explaining the observed patterns. A rich quantification toolbox was built in the last years for modelling some insect societies and can be of valuable use to shed light on many dynamical aspects of free software development.

**Acknowledgments**

We thank Kevin Crowston and James Howison for making publicly available their software data. This work has been supported by grants BFM2001-2154 and by the EU within the 6th Framework Program under contract 001907 (DELIS) and by the Santa Fe Institute.

**The Authors**

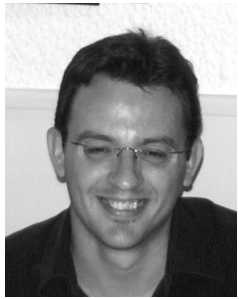

**Sergi Valverde,** is a researcher in the ICREA-Complex Systems Lab at the University Pompeu Fabra, Barcelona. He received the degree in Computer Science from Polytechnic University of Catalonia (UPC) in 1999. He was a software engineer for 10 years in the videogame and the publishing industries. His research focuses on complex networks and biologically-based modelling of artificial systems, including the Internet and software systems. He is currently involved in three European Projects (DELIS, PACE and ECagents). Contact him at the University Pompeu Fabra, Barcelona, Spain; e-mail: svalverde@imim.es; www: http://complex.upf.es/~sergi.

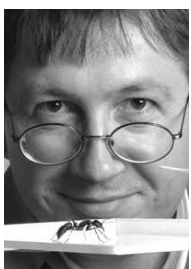

**Guy Theraulaz,** is a senior research fellow at the Centre National de la Recherche Scientifique in Toulouse, where he heads the research group on Collective Behaviours in Animal Societies (CBAS group) at the Research Center on Animal Cognition. He received a Ph.D. in neurosciences and ethology from the Provence University, (Marseille, France). He is currently working on various aspects of collective decision-making and building behaviour in social insects, collective displacements in fish schools and human pedestrians and distributed adaptive algorithms inspired by social insects. In 1996 he was awarded by the bronze medal of the CNRS for his work on Swarm Intelligence. Contact him at the Paul Sabatier University,Toulouse, France; e-mail: theraula@cict.fr; http://cognition.ups-tlse.fr/~theraulaz/.

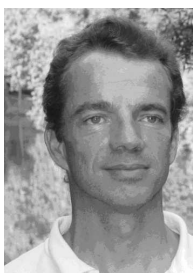

**Jacques Gautrais,** is currently researcher in the EMCC group at the Research Center on Animal Cognition, University Paul Sabatier (Toulouse, France). He is an Engineer in Biology from U.T. Compiègne in 1991 and received the Ph D degree in Cognitive Sciences from the EHESS in 1997. His research focuses on modelling self-organised collective coordination in animals, e.g. schooling, behavioural synchronization and unsupervised building. Contact him at the UPS, Toulouse, France; e-mail: gautrais@cict.fr.

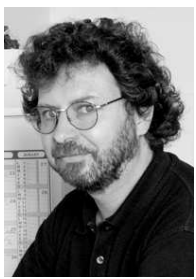

**Vincent Fourcassié,** earns a PhD from the University of Toulouse. He is currently a Research Associate with the French Centre National de la Recherche Scientifique (CNRS), and is working at the Research Center on Animal Cognition, Paul Sabatier University in Toulouse. His research interests are orientation, decision making in animal societies. Contact him at the Paul Sabatier University, Toulouse, France; e-mail: fourcass@cict.fr.

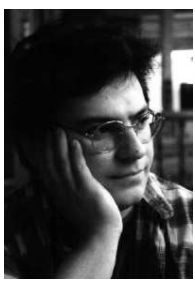

**Ricard V. Solé,** is currently research professor at the Universitat Pompeu Fabra, in Barcelona, where he is the head the ICREA-Complex Systems Lab. He received the degree in Biology (1986) and in Physics (1988) from the University of Barcelona, and the Ph D degree in Physics from Polytechnic University of Catalonia (UPC). He is also external professor at the Santa Fe Institute and senior member of the NASA-associate Astrobiology Center, in Madrid. Contact him at the University Pompeu Fabra, Barcelona, Spain; e-mail: ricard.sole@upf.edu; www: http://complex.upf.es/~ricard.

**Bibliography**

[1] R. V. Solé, R. Ferrer-Cancho, J. M. Montoya and S. Valverde, Selection, "Tinkering and Emergence in Complex Networks", Complexity, 8 (1), 2002.

[2] R. V. Solé and B. Goodwin, Signs of Life: How Complexity Pervades Biology, Basic Books, 2001.

[3] S. Wasserman and K. Faust, Social Network Analysis, Cambridge Univ. Press, New York, 1994.

[4] L. Pardi, Ricerche sui Polistini VII. "La 'dominazione' e il ciclo ovario annuale in Polistes gallicus (L.)", Boll. Ist. Entom. Univ. Bologna, 15, 25-84, 1946.

[5] M. S. Granovetter, "The Strengh of Weak Ties", American Journal of Sociology 78 (6), 1360-1380. 1973.

[6] G. Theraulaz, E. Bonabeau, J.-L. Deneubourg, "Self-organization of Hierarchies in Animal Societies", J. Theor. Biol.,174, 313-323, 1995.

[7] E. S. Raymond, "The cathedral and the Bazaar", First Monday, 3 (3), 1998.

[8] A. Barrat, M. Barthelemy, R. Pastor-Satorras, and A. Vespignani, "The Architecture of Complex Weighted Networks", PNAS, vol. 101, no. 11., 3747-3752, 2004.

[9] K. Crowston and J. Howison, "The Social Structure of Free and Open Source Software Development", First Monday, 10 (2), 2005.

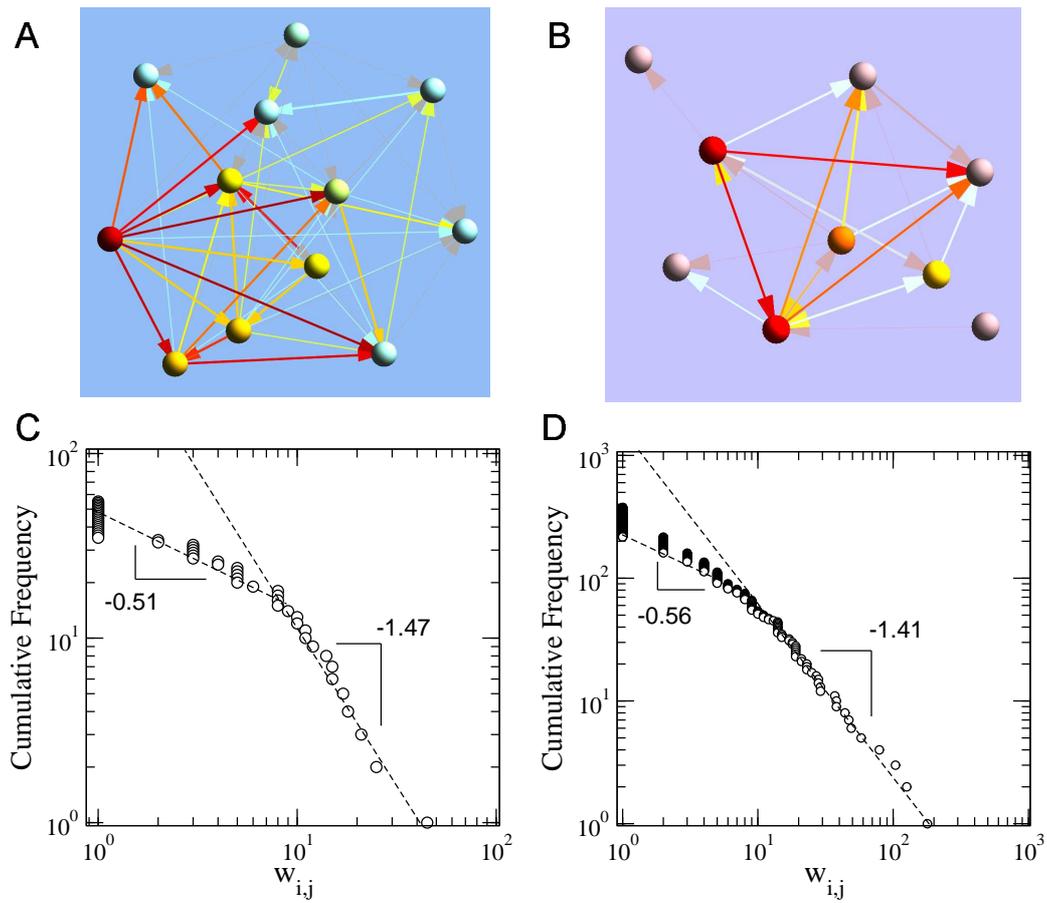

Fig.1: Heterogeneous interaction in wasp experiments and small software communities. (A) Network of hierarchical wasp interaction measured in a colony with 13 wasp members. (B) Social network of e-mail exchanges between developers in a small software community. (C) Cumulative distribution $P_>(w_{i,j})$ for a single wasp experiment. (D) Cumulative distribution $P_>(w_{i,j})$ in an ensemble of 12 small software communities. The tail of this distribution fits an scaling law $P_>(w_{i,j}) \sim w_{i,j}^{-\gamma+1}$ with $\gamma \approx 2.41$.

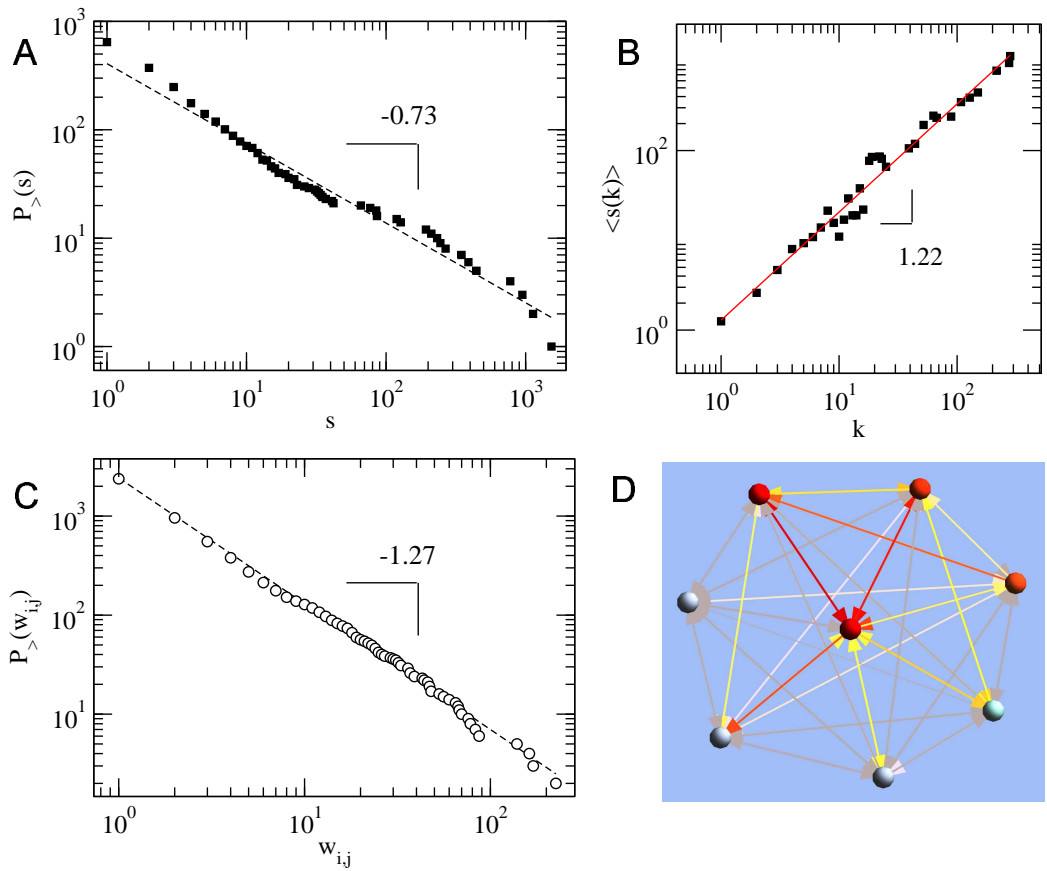

Fig. 2: Analysis of large software community. (A) Cumulative distribution for the strength $P_>(s)$ measured in the Python community. The line denote the least squares power-law fitting $P_>(s) \sim s^{-\alpha+1}$ with exponent $\alpha = 1.73$. (B) Average strength scales with out-degree with exponent 1.22. (C) The cumulative link weight distribution can be approximated by a scaling-law $P_>(w_{i,j}) \sim w_{i,j}^{-\gamma+1}$ with exponent with $\gamma \approx 2.27$. (D) Subgraph of e-mail communication between strongest developers (i.e., having $s > 200$ messages) in the Python community. Only links between core members are shown. Warmer nodes and links represent stronger developers and frequent e-mail communications, respectively.

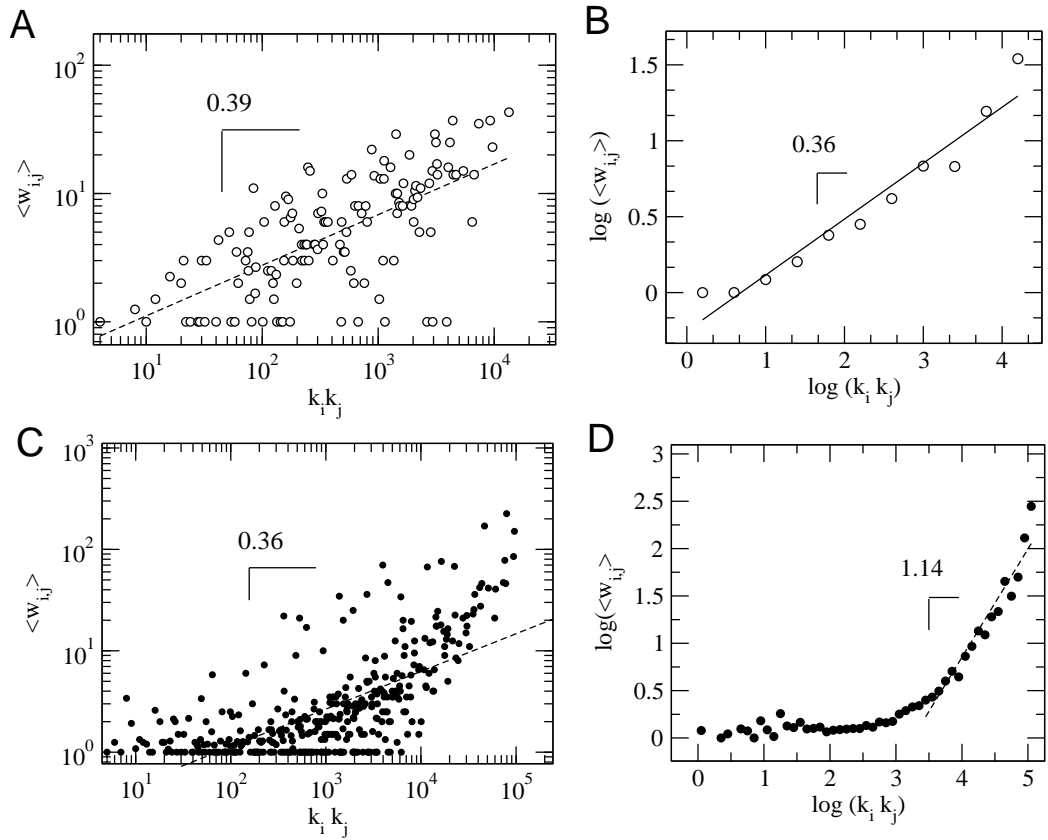

Fig. 3: Dependence of link weight $w_{i,j}$ with the product of out-degrees $k_i k_j$ measured in an ensemble of small wasp colonies and a large software community. (A) Least squares fitting of $<w_{i,j}> \sim (k_i k_j)^\theta$ in the ensemble of 5 wasp stabilized patterns yields an exponent $\theta \approx 0.39$. (B) Same as in (A) but with logarithmically binned data to reduce fluctuations. The observed scaling exponent 0.36 is consistent with (A). In (C) we show how the average link weight correlates with $k_i k_j$ in the Python project. The straight line with slope 0.36 is the exponent for the power-law fitting of the whole dataset, which shows fluctuations. (D) Same as in (C) but with logarithmically binned data. We observe an almost flat behaviour followed by a persistent deviation. The dependence of average link weight with large out-degree product fits the exponent $\theta \approx 1.14$ and thus indicating strong correlations between highly connected nodes.

BOX: **Description of wasp experiments**

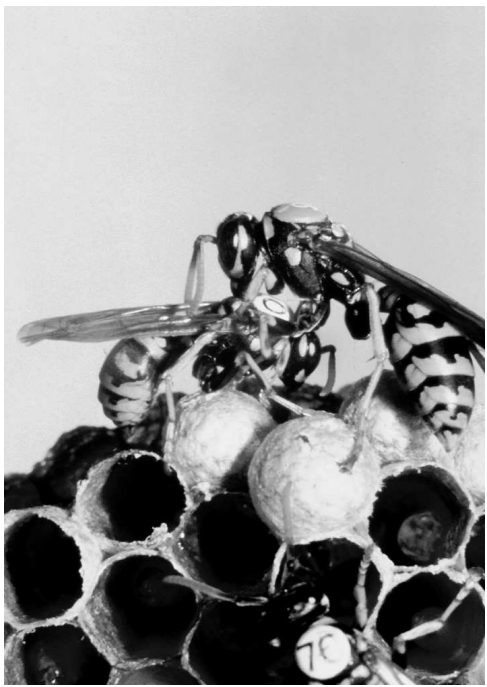

Fig.: Hierarchical interaction between wasps.

In order to uncover the network structure of wasp colonies, a set of experiments were carried out on colonies *Polistes dominulus* being the number of wasps artificially maintained constant and equal to $N_w = 13$. In this primitively eusocial species, individual behavior is rather flexible. Individuals tend to adopt specialized roles that are determined by social interactions taking place among individuals. Among these, hierarchical interactions play a crucial role and take place when two wasps encounter, thus being the result of multiple exchanges. In these encounters, the pair exhibits a dominant versus submissive role (see figure). The dominance relationship between pairs of individuals in a colony is always stable and the whole set of pair-relationships within a colony, forms a more or less linear hierarchy that may occasionally contain loops. Each wasp can therefore be allotted to a particular hierarchical rank depending on the number of individuals to which it is subordinate in the colony. In this particular species an individual's hierarchical rank generally coincides with its order of birth (and thus aging effects are at work).

The experiments involved the removal of top rank individuals (also called the α-individual) and studying the subsequent re-organization of the colony distribution of activities. The study was carried out in parallel on two colonies. Every week, the wasp occupying the top of the hierarchy was removed. This period of time appears to be sufficient to allow the new hierarchical pattern to become stabilized. The observation period lasted for 38 days (5 weeks). The data were recorded during 2h observation sessions every day. The wasp's behavior was observed visually and data recorded on a micro computer the keyboard of which was re-designed for behavioral coding. Any social contacts occurring between pairs of individuals were recorded and the weight of each pair of interactions was measured in terms of the number of (directed) contacts.